# Spheroid Model for Molecular Packing in Crystalline Phase

Weihao Wang,[1] Zhenghong Chen,[1] Yang Gao,[2] Yang Jiao,[3] Shaodong Zhang[1,*]

[1]School of Chemistry and Chemical Engineering, Shanghai Jiao Tong University, Shanghai 200240, China

[2]School of Mathematical Sciences, Shanghai Jiao Tong University, Shanghai 200240, China

[3]Materials Science and Engineering, Arizona State University, Tempe, AZ 85287

**Abstract:** Dense packing of particles has provided important models to study the structure of matter in various systems such as liquid, glassy and crystalline phase, etc. The simplest sphere packing models are able to represent and capture salient properties of the building blocks for covalent, metallic and ionic crystals; it however becomes insufficient to reflect the broken symmetry of the commonly anisotropic molecules in complex molecular crystals. Here we develop spheroid models with the minimal degree of anisotropy, which serve as a simple geometrical representation for a rich spectrum of molecules——including both isotropic and anisotropic, convex and concave ones——in crystalline phases. Our models are determined via an "inverse packing" approach: given a molecular crystal, an optimal spheroid model is constructed using a "contact diagram", which depicts packing relationship between neighboring molecules within the crystal. The spheroid models are capable of accurately capturing the broken symmetry and characterizing the equivalent volume of molecules in the crystalline phases. Our model also allows to retrieve such molecular information from poor-quality crystal X-ray diffraction data that otherwise would be simply discarded.

In the crystalline phase the packing arrangement of individual building blocks (BD), including atoms, ions, molecules[1,2], nano-particles[3], and colloidal particles[4], to name but a few, plays a pivotal role in determining the properties of the matters that they form, which is therefore of great interest in the modern era of physical, chemical and materials sciences[5,6]. Taking covalent, metallic and ionic crystals for instance, their structures have been successfully described by sphere packing modeling, in which their corresponding BDs are represented by the simplest and fully isotropic geometrical shape[5], capturing salient fundamental properties of the BDs (such as symmetry). On the other hand, the sphere models are generally insufficient to describe complex molecular crystals, as the isotropic shape cannot reflect the broken symmetry of the molecules (illustrated in Fig. S1). Geometric representations with additional degrees of freedom are thus required for complex molecules vis-à-vis the molecular packing in the crystalline phases.

Recently, dense crystalline packings of a variety of anisotropic particles including ellipsoids[7], tetrahedra[8], superballs[9,10] and polyhedra[11,12] have been investigated, as such nonspherical particles provide improved and more realistic representations of complex anisotropic BDs for a variety of condensed matters[13,14]. Among these shapes, spheroids are a family of symmetry-breaking shapes with *minimal degree of anisotropy*, yet they have demonstrated a rich spectrum of packing behaviors, and were employed in the study of Frenkel-Mulder contact diagram[15,16], nematic phase transition[17,18] and quasi-crystals[19]. These intriguing studies imply that spheroids would also provide an effective representation for the "equivalent volume" of anisotropic complex BDs in the crystalline phase, i.e., the volume of the occupied space of each BD that is inaccessible by others.

We herein report on a simple geometrical representation for complex molecules in crystalline phases, which is based on spheroid particles. Our models are determined via an "inverse packing" approach: Unlike traditional packing problems that focus on finding the optimal packing arrangement for a given particle shape, our "inverse packing" approach identifies the optimal spheroid shape (defined by the two semi-axis $R_1$ and $R_2$ in Fig. S1) that represents the molecules, when given their crystalline packing structure. To achieve this goal, we devise a contact diagram which depicts the packing relationship (overlapping, contact, non-touching) of all representative spheroid pairs derived by the symmetry operations within a molecular crystal.

By investigating distinct types of molecular crystals, we demonstrate the spheroid model can successfully capture the broken symmetry and key features of the molecules in their crystalline phases. Our work also provides a new paradigm that complex BDs can be represented by simple shapes allowing overlapping, instead of sticking to hard-particles with increasing shape complexity. Although omitting many complex molecular details, the spheroid models are capable of accurately characterizing the equivalent volume of molecules in the crystalline phases. This is to contrast the widely used Van der Waals (VdW) volume[20,21], which typically contains redundant structural information and requires nontrivial computation even for relatively simple molecules (such as $CH_4$). Moreover, our models are also effective even with crystals with poor quality (i.e., those reconstructed from noisy data), from which valuable packing information can be extracted as useful input for further optimization of molecular design.

## Results

**Procedure for inverse packing problem.** We first describe the inverse packing procedure: the centroid of a molecule is represented with the center of a spheroid, which is placed on a site of the molecular crystal lattice. The principal direction of the spheroid (i.e., direction of the axis of resolution of the spheroid) corresponds to that of the molecule, which is typically associated with the direction of molecular dipole moment. For nonpolar molecules, their principal symmetry axis is used to align with the principal direction of the spheroid. Once the position and direction of spheroids are determined, the osculation of two adjacent spheroids can be described using the relationship between their respective semi-axis $R_i$ and $R_j$ (see Fig. S2 for illustration). This $R_i/R_j$ correlation is derived according to the Perram and Wertheim contact function[22], which corresponds to a curve in the contact diagram, as illustrated with Fig. 1(c) for example.

A contact diagram is constructed with a series of contact curves, which depict whether a pair of spheroids with a specific semi-axis ratio $R_i/R_j$ overlap, contact or mutually separated (illustrated in detail in Fig. S3). This in turn allows us to determine the dimension of spheroids that represents the equivalent volume of the molecules within the crystalline phase, as described in detail below.

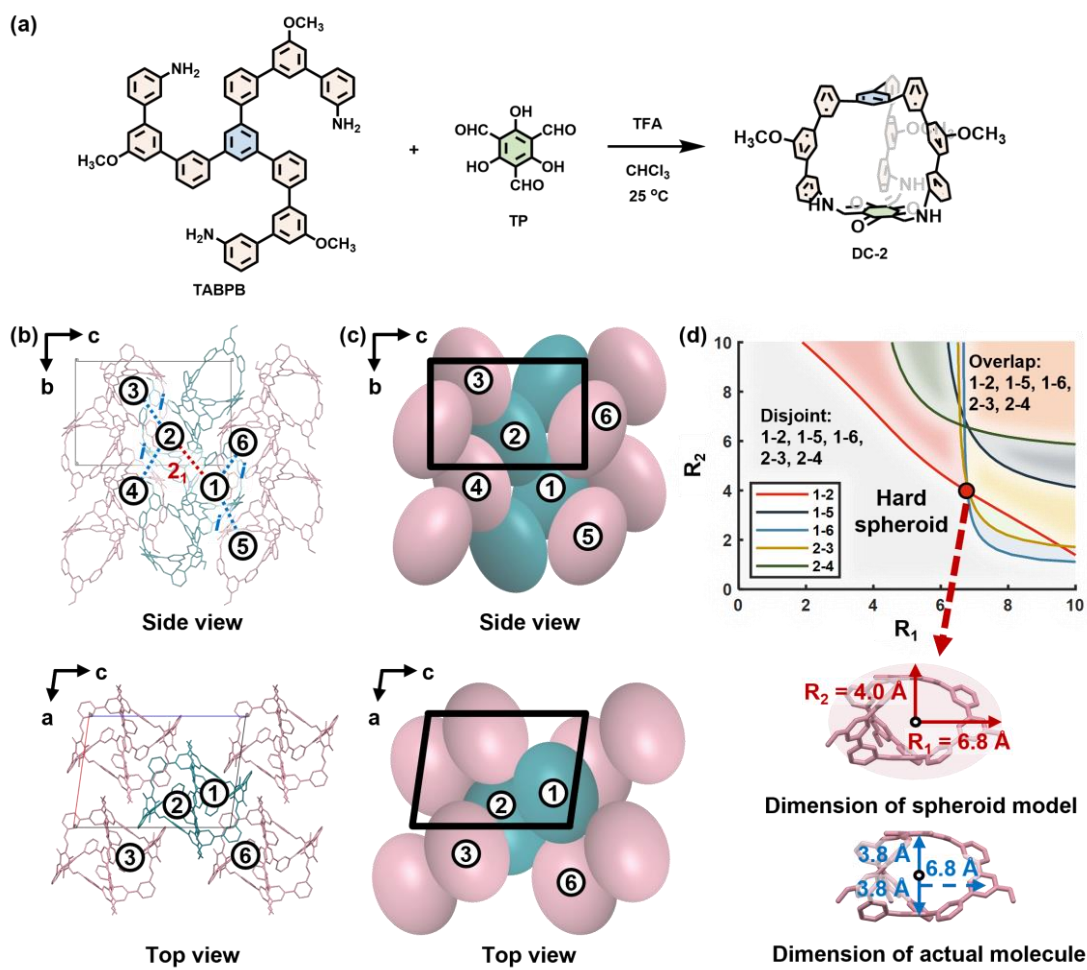

**Fig. 1** Spheroid packing model for a crystal of cage-like molecules with space group $P2_1/n$. (a) Synthesis of molecule cage. (b) Side and top views of crystal. (c) Side and top view of simulated spheroid packing with magenta and cyan colors used for easy inspection. (d) Computed contact diagram of hard spheroid packing. Curves with different colors correspond to contacts between different pairs of molecules. Simulated hard spheroid shows similar dimension with that of the actual molecule.

**Hard spheroid model applied to molecular crystal.** As a class of novel molecules, cages-like compounds have attracted considerable attention for the recent decades[23–31]. By virtue of their rich geometric diversity, molecular cages can be used as promising BDs for the search of novel supramolecular materials which are hardly accessible by conventional molecules[32]. As a proof of concept, cage molecules with rigid and well-defined 3D structure are first selected to demonstrate our spheroid model for molecular crystals. The spheroid packing is constructed by considering symmetry constraints imposed by the crystal lattice, which is first illustrated with a crystal formed by molecular cage that we synthesized according our previous protocol (Fig. 1)[33]. The molecular cage is synthesized by the reaction between 2,4,6-trihydroxybenzene-1,3,5-tricarbaldehyde (TP) and a triamino-functionalized TABPB, as illustrated in Fig. 1(a). Their detailed synthetic procedures, NMR and MALDI-TOF analyses are given in the Supplemental Information.

In this crystal with space group of $P2_1/n$, a molecule is related to its neighbors through symmetry operations of inversion ($i$), screw (with $2_1$ axis) and glide, with latter a product of the first two operations. Therefore, we only need to consider inversion and screw symmetry constraints to construct a local cluster of spheroids, from which the "contact diagram" will be derived. To explain the contact imposed by two-fold screw operations, two adjacent spheroids labeled as 1 and 2 on screw axis are first selected. Each of the two spheroids further osculate two neighboring spheroids with inversion constraint, which are labelled from 3 to 6, respectively. As a result, only a local cluster of six spheroids are required to elucidate all contact relationships in such case, as illustrated in Fig. 1(c).

Fig. 1(b) reveals the crystal structure viewed along crystallographic $a$ and $b$ axes, and the packing of the corresponding spheroidal model is shown in Fig. 1(c). Taking the contact of spheroids 1 and 2 for instance, with their orientations and center positions fixed, $R_1$ and $R_2$ cannot vary independently while maintain the contact——they need vary coherently, leading to a correlation between $R_1$ and $R_2$ depicted by the red line in the contact diagram in Fig. 1(d). The two spheroids are disjoint when the coordinate ($R_1$, $R_2$) is below of the line, while they are intercalated when the coordinate is above. By iterating the $R_1/R_2$ relationship with other spheroid pairs, all five lines reflecting their osculation are derived. As a result, the contact diagram is divided into various regions that correspond to different packing patterns. For example, the grey region in the bottom left area shows that all spheroids are disjoint, and the orange part on the top right corner indicates that all spheroids overlap with each other.

For hard spheroids overlap is prohibited, and two arbitrary spheroids can only be tangent or disjoint to each other. As the leftmost intersection (red circle in Fig. 1(d)) reaches the largest contact number while guarantees no overlap of spheroids, this coordinate is therefore deemed as the dimension of the hard spheroid. Accordingly, the polar and equatorial radii of the spheroid are calculated to be 4.0 and 6.8 Å, respectively. This calculated dimension is almost identical to that of the actual molecule with a height of 7.6 Å and a radius of 6.8 Å.

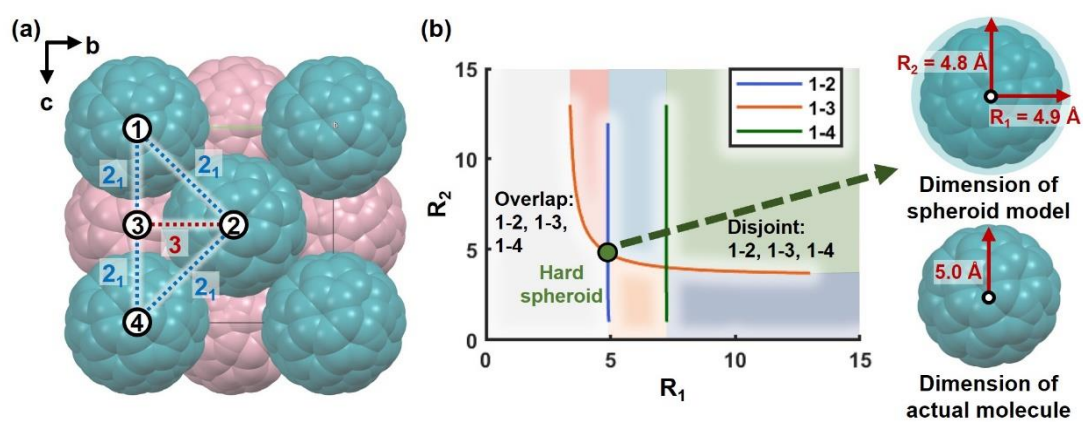

**Fig. 2** Spheroid packing in fullerene crystal with space group *Pa*3. (a) Crystal structure viewed along *a* axis. Adjacent molecules are related through different symmetry operations. (b) Computed contact diagram of spheroid packing. The simulated shape and dimension are close to those of a fullerene molecule.

This spheroid model can also be applied to the molecular crystals formed by (quasi)isotropic molecules, which is exemplified with landmark molecule fullerene $C_{60}$ with large void[34]. The VdW volume only counts the sum of the occupied space of all atoms but excludes the cavity, and thus, cannot accurately characterizes the equivalent volume of such type of molecules in dense crystalline packing. As illustrated in Fig. 2, the molecular crystal (space group *Pa*3) is constrained by screw operation (with $2_1$ axis) and rotation (3-fold axis). Therefore, only four spheroids are required to describe all the contact relationships in Fig. 2(a), which correspond to three contact lines of 1-2, 1-3, and 1-4 spheroid pairs. Similarly, the coordinate of the leftmost intersection (in red circle) refers to the two semi-axes of the hard spheroid, *i.e.*, 4.9 and 4.8 Å, respectively. These values are in good accordance with the radius of a fullerene in the crystalline phase. We also verified the applicability and generality of the spheroid model to a variety of randomly selected molecules with regular shape, which will be discussed afterwards.

**Soft spheroid model for dovetailed molecules.** We further examined a ubiquitous type of molecules that dovetail with each other during crystallization. This is showcased with our recently reported twin-cavity cage[35]. Fig. 3(a) manifests the molecular packing viewed along $c$ axis in the crystal (space group $P3_221$). As their packing is imposed by symmetry operations of screw (with $3_2$ and $2_1$ axes) and rotation (2-fold axis), five different spheroids labeled from 1 to 5 are selected accordingly.

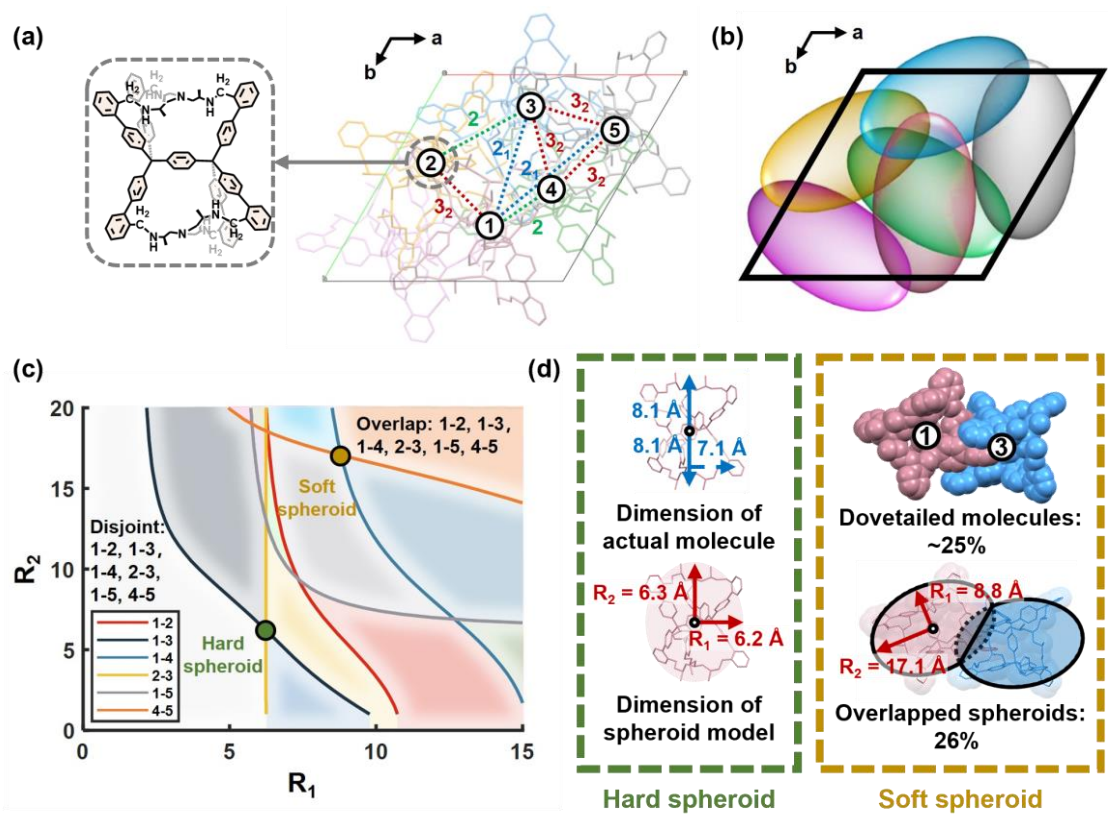

**Fig. 3** Spheroid packing in a crystal (space group $P3_221$) formed by dovetailing twin-cavity cages. (a) Crystal structure viewed along $c$ axis. Adjacent molecules are constrained by symmetry operations. (b) Simulated soft spheroid packing; (c) Computed contact diagram of spheroid packing. The innermost node represents hard spheroid, while outermost node represents soft spheroid. (d) Calculated hard spheroid is smaller than actual molecule, and soft spheroid reflects the dovetail between molecules based on the overlap percentage.

Similar to the aforementioned analysis, the hard spheroid corresponds to the intersection of dark blue and yellow lines in the contact diagram in Fig. 3(c), corresponding to a spheroid with polar and equatorial radii of 6.3 and 6.2 Å, respectively. However, compared with the actual dimension of the molecule, the hard spheroid obviously contracts as pictured in Fig. 3(d). This is because the compounds are indeed interpenetrated in the crystalline phase, and the use of hard spheroid model unavoidably excludes the dovetailed part, which therefore underestimates the equivalent volume of such molecules.

To better delineate the interpenetration of this class of molecules, we introduced a supplementary soft (i.e., overlapping) spheroid model[36]. Unlike hard spheroids, soft spheroids are elastic that allow deformation, so that all spheroids of interest are either contacting or overlapping. It means their disconnection is avoided, and the soft spheroid packing in this crystal is displayed in Fig. 3(b). Accordingly, the outermost node is the critical point for reaching the largest contact number, which is chosen as the coordinate for presenting the dimension for the soft spheroid (Fig. 3(c)). Notably, if there are more than one innermost or outermost nodes, the one with the largest value of $R_1^2 R_2$ is chosen. This is because this value corresponds to the largest spheroid volume, complying with the highest packing fraction in the crystalline phase. Additionally, overlap of spheroids is forbidden by translation operation, as it is not consistent with the dovetailing of the particles. The largest polar and equatorial radii of the soft spheroid are therefore calculated to be 17.1 and 8.8 Å respectively; the interpenetration percentage is about 26%, while the dovetail of actual molecules is around 25% as shown in Fig. 3(d). Indeed, the hard and soft spheroids respectively determine the lower and upper limits of the equivalent dimension for a specific molecule in the crystalline phase, and the most suitable shape model should depend on the convexity/concavity of the molecule.

**Comparison between hard spheroid model and space-filling model.** We subsequently compared our hard spheroid model with the conventional space-filling VdW model, both of which were used to probe the equivalent volume of molecules in the crystalline phase. To this end, seventeen extra molecules including peptides, cholesterols, triptycene, macrocycles and other small molecules were analyzed, with their contact diagrams shown in Fig. S4–S7, and their VdW volumes were calculated with multifunctional program Multiwfn[37]. The comparison of the calculated VdW volume and hard spheroid volume is shown in Fig. 4. The data points close to the yellow line indicate similar values of volume in the two models. Different colors of the data points indicate different types of molecules: general molecules with regular shape are shown in blue, red dots represent compounds with cavity, and purple symbols stand for some seriously dovetailing molecules.

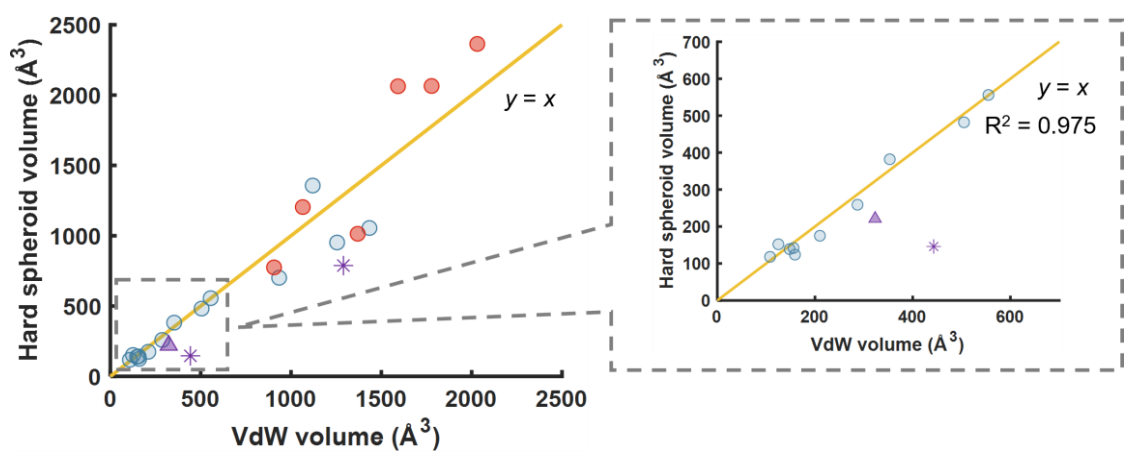

**Fig. 4** Comparison of simulated hard spheroid volume and Van der Waals (VdW) volume of a wide spectrum of complex molecules. Red dots represent molecules with cavity, purple asterisks represent situations for severely dovetailed molecules, and blue dots represent other molecules. Enlarged is the simulated hard spheroid volume of small molecules showing good linear relationship with their VdW volume.

Focusing on the enlarged area in Fig. 4, we can find that the hard spheroid volume is fairly close to the VdW volume for relatively small and regular molecules. This is because these molecules are compact and don't interpenetrate in their crystalline packings, and both space-filling model and spheroid model are applicable. However, for porous molecules, their intrinsic cavity is taken into account by the hard spheroid model, whose dimension is thus larger than the corresponding VdW volume that does not consider the cavity. Our spheroid model therefore can better describe the equivalent volume of such molecules. Moreover, for the molecules with concave structure and/or flexible fragments, they are often prone to dovetail with each other during molecular packing. This information, particularly the penetration percentage (*vide supra*), can hardly be unraveled by the VdW model. In hard spheroid model, overlap between spheroids are forbidden, and the calculation of spheroid volume will exclude the dovetailed part, causing an underestimation of the equivalent volume. To address this problem, the soft spheroid model is proposed, which can depict the dovetail through interpenetration percentage between the spheroids.

**Table 1** The calculated hard spheroid volume and VdW volume of seventeen different molecules collected from Cambridge Crystallographic Data Centre (CCDC).

| Compound | CCDC number | Space group | Hard spheroid volume (Å$^3$) | VdW volume (Å$^3$) |
|---|---|---|---|---|
| Small molecule | 202500 | $P\bar{1}$ | 110 | 190 |
| Small molecule | 1133112 | $Cc$ | 139 | 148 |
| Triethylenediamine | 1269548 | $P6_3/m$ | 118 | 108 |
| Ferrocene | 1154857 | $P2_1/a$ | 152 | 125 |
| Cortisone | 2010417 | $P2_12_12_1$ | 221 | 323 |
| Cholic acid | 1116202 | $P2_1$ | 396 | 381 |
| Helicene | 1521681 | $P2_12_12_1$ | 259 | 287 |
| Triptycene | 1275698 | $P2_12_12_1$ | 175 | 210 |
| Corannulene derivative | 1567460 | $P3c1$ | 2424 | 2439 |
| Vitamin B12 | 1944201 | $P2_12_12_1$ | 1357 | 1120 |
| Chain-like molecule | 257085 | $P\bar{1}$ | 1055 | 1434 |
| Peptides | 1190276 | $P2_1$ | 146 | 443 |
| Macrocycle | 1822570 | $P2_1/c$ | 556 | 555 |
| Macrocycle | 1950613 | $R\bar{3}$ | 952 | 1255 |
| Cage | 1453937 | $R\bar{3}c$ | 2063 | 1592 |
| Cage | 1566755 | $C2/c$ | 2364 | 2031 |
| $C_{76}Cl_{16}$ | 650718 | $Pbca$ | 701 | 934 |

**Spheroid model applied to poor-quality crystal.** The proposed spheroid model can also be applied to probe the molecular packing of crystals with incomplete structural refinement, which often result from poor quality of crystal X-ray diffraction data. These data are conventionally considered invalid and are therefore simply discarded. Illustrated in Fig. 5 is the contact diagram derived from the poorly resolved crystalline phase of the molecules used in Fig. 1. This rough structure was obtained from original crystal data without further refinements, with a *R*-factor of *ca*. 39%[38]. The semi-axes of the simulated spheroid were determined to be 6.8 Å and 3.7 Å, respectively, which are close to the spheroid dimension calculated from the precise structure. Our spheroid model therefore can provide a tool for extracting valuable information of molecular packing in crystals of poor quality, which would be useful input for the optimization of molecular design.

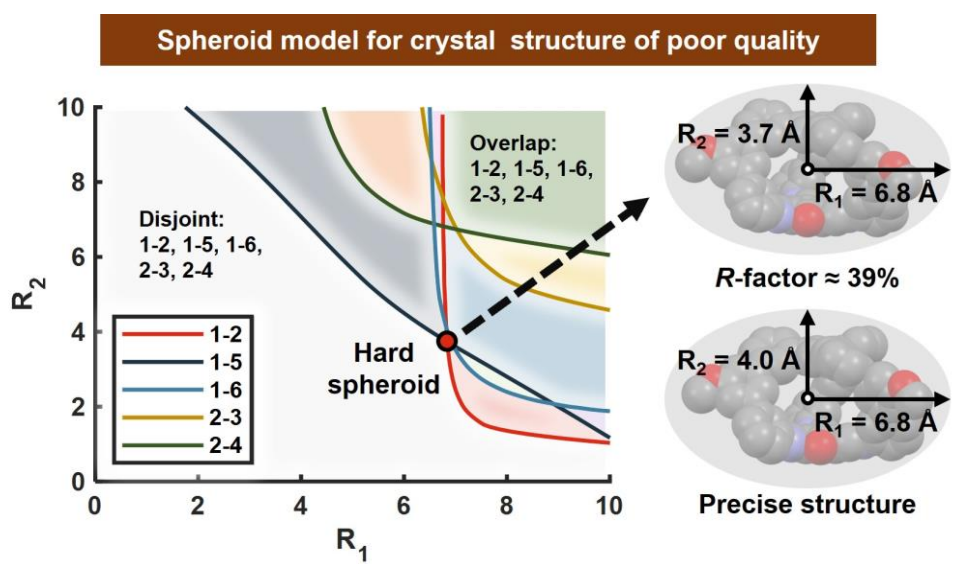

**Fig. 5** Spheroid model applied in poor-quality crystals. The contact diagram is derived from the rough structure (*R*-factor around 39%) and its simulated spheroid is close to the spheroid calculated from precise structure.

## Discussion

Just like the coarse-grained model that omits the detailed complexity yet maintains the main peculiarities of a molecule, our proposed spheroid model employs simple geometrical shapes and focuses on the salient anisotropic features of complex molecules. We have demonstrated that the relationship between molecule anisotropy and its crystal packing can be better captured by our model. In contrast to space-filling models, our model better reflects the equivalent volumes of certain molecules with exotic shapes. It also provides information of dovetail between adjacent molecules within the crystalline phase. We expect our model can be readily generalized to more complex molecular crystals by increasing the degrees of freedom of the fundamental geometric representation, e.g., from spheroid to generic ellipsoid or other shapes, which we will explore in our future work.

## Methods

**Computational methods.** We take Perram-Wertheim contact function to probe the spatial relationship between spheroids. This function reveals the correlation between spheroid position, orientation, dimension and osculation. First of all, the position and direction information of a spheroid can be determined by specific crystal lattice, which is provided by the Cambridge Structural Database (CSD). Next, we assume two spheroids osculate with each other, which is realized by adjusting $R_1$ and $R_2$ to ensure that the value of contact function is close to one. Every osculation of a spheroid pair will therefore correspond to a curve in the contact diagram, showing the variation of $R_1$ and $R_2$. As only limited symmetry operations are present in a crystal, the sort of packing relationship (overlapping, contact, non-touching) between adjacent spheroids is also finite. Therefore, we only need to analyze those representative spheroids pairs that contain all contact relationships in a crystal. As a result, the contact diagram consists of several curves that reflect the osculation of different spheroid pairs, and these curves divide the diagram into various sections representing different packing relationships of spheroids.

**Hard and soft spheroid model.** The selected hard and soft spheroid, prohibiting or allowing overlap, respectively correspond to the leftmost and rightmost intersection. Therefore, the volume of hard spheroid can be easily determined by MATLAB, which is then compared with its VdW volume calculated by Multiwfn. Additionally, the interpenetration percentage of soft spheroids, defined as the quotient of overlapped volume and soft spheroid volume, is derived by Monte Carlo algorithm and is used to describe the dovetail between molecules with concave structure or flexible fragments.

## Acknowledgments

We thank financial support from the Science and Technology Commission of Shanghai Municipality (21JC1401700) and the National Natural Science Foundation of China (21890733, 22071153, 22271187). The crystallographic experiments were conducted with beam line BL17B1 supported by Shanghai Synchrotron Radiation Facility.

## Author contributions

W.W. and S.Z. designed this work. W.W. and Y.G. built the mathematical model of spheroid packing. W.W. and Z.C. collected and analyzed the data. W.W., Y.J. and S.Z. discussed the results and prepared the paper.

## Competing interests

The authors declare no conflict of interest.